\documentclass[preprint,letterpaper,showpacs]{revtex4}
\usepackage{graphicx}
\def \uchi {\underline{\chi }}
\def \ugamma {\underline{\Gamma }}

\def \nc {n_{\chi }}
\def \Seff {S_{\rm eff}}
\def \qsg {q_{\rm SGL}}
\def \qfm {q_{\rm FM}}
\def \om {\omega_m}

\def \vS {{\bf S}}

\def \nc {n_{\chi }}

\def \vq {{\bf q}}

\def \om {\omega_m}

\def \hm {{\bf m}}

\def \TC {T_{\rm C}}

\def \TM {T_{\rm SGL}}
\def \TR {T_{\rm SGL}}
\def \Tsg {T_{\rm SGL}}
\def \Jeff {J_{\rm eff}}

\def \E2min {E_2^{({\rm min })}}

\def \bs {{\bf \sigma }}
\def \vS {{\bf S}}

\def \vs {{\bf s}}

\def \JH {J_{\rm H}}

\begin{document}

\title{Magnetic Susceptibility of the Double-Exchange Model in the RKKY Limit}
\author{Randy S. Fishman}
\affiliation{Materials Science and Technology Division, Oak Ridge National Laboratory, Oak Ridge, TN 37831-6032}

\begin{abstract}

The magnetic susceptibility and Edwards-Anderson order parameter $q$ of the spin-glass-like (SGL) 
phase of the double-exchange model are evaluated in the weak-coupling or RKKY limit.   
Dynamical mean-field theory is used to show that $q=M(T/\Tsg )^2$, where $M$ is the classical 
Brillouin function and $\Tsg $ is the SGL transition temperature.  The correlation length of the SGL 
phase is determined by a correlation parameter $Q$ that maximizes $\Tsg $ and minimizes the 
free energy.  The magnetic susceptibility has a cusp at $\Tsg $ and reaches a nonzero value as 
$T\rightarrow 0$.  

\end{abstract}

\maketitle

One of the most important models of itinerant systems, the double-exchange (DE) model
is believed to describe many physical systems ranging from the manganites \cite{mang} to dilute magnetic
semiconductors \cite{dms}.  In the weak-coupling limit, the DE model becomes equivalent \cite{van:62}   
to a RKKY model with competing antiferromagnetic (AF) and ferromagnetic (FM) Heisenberg 
interactions between classical spins at every site.  Recent work \cite{fis:06} has revealed 
that these competing interactions can stabilize a phase with short-range but not long-range 
magnetic order.   
We now show that the magnetic susceptibility of this spin-glass-like (SGL) phase has a cusp at 
$\Tsg $, marking the onset of short-range order, and reaches a nonzero value as 
$T\rightarrow 0$.   The Edwards-Anderson (EA) order parameter $q$ \cite{edw:75,sher:75} 
of the SGL phase is identical to the square of the classical Brillouin function $M(T/\Tsg )$.

The DE model contains a kinetic term that describes the hopping of electrons between 
neighboring sites and a potential term that aligns the electronic spins with the classical local 
moments at every site.   There is no quenched disorder in the DE Hamiltonian
\begin{equation}
\label{ham}
H=-t\sum_{\langle i,j \rangle }\Bigl( c^{\dagger }_{i\alpha }c^{\, }_{j\alpha }
+c^{\dagger }_{j\alpha }c^{\, }_{i\alpha } \Bigr) -2 \JH \sum_i \vs_i \cdot \vS_i,
\end{equation}
where $c^{\dagger }_{i\alpha }$ and $c_{i\alpha }$ are the creation and destruction operators 
for an electron with spin $\alpha $ at site $i$, 
$\vs_i =(1/2) c^{\dagger }_{i\alpha } \bs_{\alpha \beta } c^{\, }_{i\beta }$
is the electronic spin, and $\vS_i=S\hm_i $ is the classical spin of the local moment.  
Repeated spin indices are summed.   When $\JH > 0$, the Hund's coupling favors the alignment of 
the local moments with the electronic spins.  Due to the electron-hole symmetry of the DE model, we shall
only consider electron concentrations $p$ between 0 and 1 carriers per site.  For small $p$, 
the hopping of electrons between neighboring sites favors the alignment of the local moments and the FM phase is 
stable.  Since electrons with parallel spins cannot hop between singly-occupied sites due to the Pauli 
exclusion principle, the AF phase is favored over the FM phase near $p=1$.  But as shown in Ref.\cite{fis:06}, the competing FM and AF interactions may actually favor a SGL phase over the ordered 
phases for small $\JH $.

Developed in the late 1980's by M\"uller-Hartmann \cite{mul:89} and Metzner and Vollhardt \cite{met:89}, 
dynamical mean-field theory (DMFT) exploits the momentum independence of the 
electronic self-energy in infinite dimensions.  Even in three dimensions, DMFT is believed to 
capture the physics of correlated systems including the narrowing of electronic bands and the 
Mott-Hubbard transition \cite{geo:96}.  Within DMFT, the local effective action on any site is 
parameterized by a Green's function that regulates the hopping of correlated electrons 
from other sites.   We will use DMFT to study the DE model on a Bethe lattice in infinite dimensions.
The bare density-of-states of a Bethe lattice with $z \gg 1$ nearest neighbors is 
$N_0(\mu )=(4/\pi W)\sqrt{1-(2\mu /W)^2}$, where $W=4\sqrt{z}t$ is the bandwidth and $\mu $ 
is the chemical potential.  Since the Bethe lattice is {\it not} translationally invariant, $\vq =0$ is the 
only well-defined wavevector \cite{fis:06}

In infinite dimensions, the high-temperature non-magnetic (NM) phases of the 
Heisenberg and DE models have a vanishing correlation length $\xi $.   
The SGL phase is a bulk solution of the DE model with some of the same characteristics as
conventional spin glasses:  a finite local magnetization and spin-spin correlations
that decay exponentially over distance \cite{bin:86}.  The SGL phase is characterized by a 
correlation parameter $Q$, defined as the average over all neighbors of $\sin^2 (\theta_i/2) $, 
where $\theta_i $ is the angle between the central spin and a neighboring spin.  Overall, the 
neighboring spins describe a cone with angle $2\arcsin (\sqrt{Q})$ around the central spin.  
The FM and AF phases have, respectively, $Q=0$ and 1.  The magnetization about every site 
decays exponentially with a correlation length $\xi = -a/\log \vert 2Q-1\vert $, where $a$ is the 
lattice constant.  Notice that $\xi /a$ diverges in the FM and AF phases but vanishes in the 
NM state obtained by setting $Q=1/2$.   Mathematically, the SGL phase was first introduced 
by Chattopadhyay {\em et al.} \cite{cha:01}, although its physical significance was not 
recognized until later.  In lower dimensions, the SGL phase evolves into the phase with incommensurate 
correlations obtained in Monte-Carlo simulations \cite{yun:98}.

The transition temperature $\TM (p,Q)$ of the SGL phase may be evaluated from coupled Green's 
function equations \cite{fis:06}.   In the weak-coupling limit $\JH S \ll W$ and $T \sim (\JH S)^2/W $, 
$\TM (p,Q)$ is implicitly given by the expression $\Tsg (p,Q) = \Jeff (p,Q)/3$ where
\begin{equation}
\label{TM}
\Jeff (p,Q) = -2(\JH S)^2 (2Q-1)T\sum_n \frac{R_n}{(z_n+R_n)^2 (z_n+2(1-Q)R_n)},
\end{equation}
with $R_n=\Bigl(-z_n +\sqrt{z_n^2-W^2/4}\, \Bigr)/2$, $z_n=i\nu_n +\mu$, and $\nu_n=(2n+1)\pi T$. 
Since $T\ll W$, the sum $T\sum_n F(\nu_n) $ is equivalent to the integral $(1/2\pi )\int dv F(v )$ and  
$\Jeff (p,Q)$ is independent of temperature.  The relation $\Tsg (p,Q)=\Jeff (p,Q)/3$ correctly reduces to the 
FM ($Q=0$) result first derived in Refs.\cite{fur:95} and \cite{aus:01} .   A similar derivation of $\Tsg (p,Q)$ 
was recently provided in Ref.\cite{kog:06}.  Of course, $\Tsg (p,Q)$ vanishes in the NM state with $Q=1/2$.   

After $\Tsg (p,Q)$ is maximized with respect to $Q$, $\TR (p)$ exceeds the 
Curie and N\'eel temperatures in the concentration range $0.26 < p < 1$.  The correlation parameter $Q$ 
changes discontinuously at $\TR (p)$ from $1/2$ in the NM phase above to a value less than or greater than 
$1/2$ in the SGL phase below.  The ground state is NM with $Q=1/2$ for a single concentration close 
to $p=0.5$.

The temperature dependence of the local SGL order parameter $M =\vert \langle \hm_i\rangle \vert $
on site $i$ may be evaluated in a local environment fixed by the correlation parameter $Q$ by integrating the
local action over the Fermion variables.  The probability for $\hm_i $ to point
at an angle $\cos \theta $ with respect to the local quantization axis is proportional to $\exp (M \Jeff \beta \cos \theta )
=\exp \bigl((3 M \cos \theta )/\tau \bigr)$, where $\tau =T/\Tsg $.
Consequently, $M$ has the solution $M (\tau ) = \coth (3M /\tau )-\tau /(3 M )$, which is just the Brillouin function
in the $S\rightarrow \infty $ or classical limit \cite{yos:91}, plotted in Fig.1.  The result for the FM order parameter is not surprising considering
the weak-coupling equivalence between the DE model and a Heisenberg model with RKKY interactions between
classical spins.  What is surprising is that the short-range order parameters of the SGL phase are identical to the 
long-range order parameters of the FM and AF phases.  For small $\tau $, 
$M(\tau ) \approx 1-\tau /3 -\tau^2/9 +\vartheta (\tau^3 )$.  

\begin{figure}
\includegraphics *[scale=0.6]{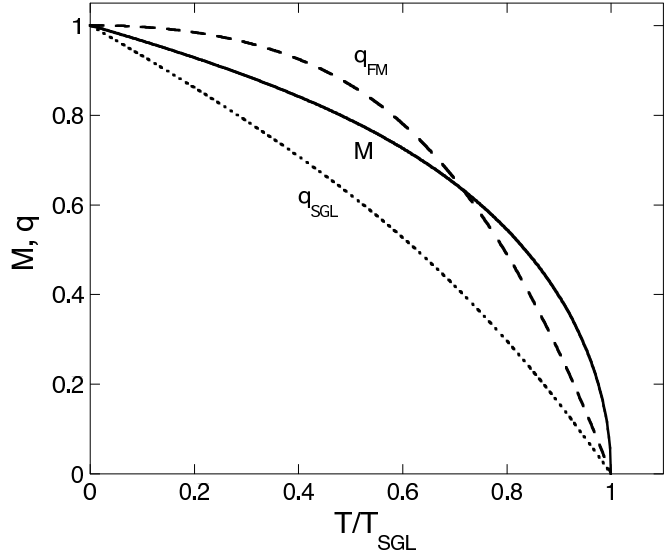}
\caption{
The Brillouin function $M(\tau )$ and EA order parameters $q(\tau )$ versus $\tau = T/\Tsg $.
}
\end{figure}

To zeroth order in $\JH S/W$, $p$ is given in terms of $\mu $ by
$p=1/2+(1/\pi )\Bigl\{ \delta \sqrt{1-\delta^2 }+\sin^{-1}\delta \Bigr\}$,
where $\delta =2\mu /W$.  Carefully accounting for the dependence of the chemical potential $\mu (p)$ on 
$\JH S/W$ for a fixed $p$, we have generalized the $T=0$ relation derived in 
Ref.\cite{fis:06} for the energy difference $\Delta E (p,Q)$ between the SGL and NM phases:   
$\Delta E(p,Q)/N =-(3/2)M(\tau )^2\,  \Tsg (p,Q)$, 
where both sides are evaluated analytically to order $(\JH S)^2/W$.   This relation is precisely the same as the 
MF result for the FM phase of a classical Heisenberg model.  By integrating the specific heat to obtain the entropy, 
we have formally constructed the free energy difference $\Delta F (p,Q)$ \cite{ent} between the SGL and NM phases.  
Because $(1/N)\partial \Delta F/\partial Q =-(3/2)M(\tau )^2 \, \partial \Tsg /\partial Q $, the free energy is minimized by the same 
correlation parameter $Q$ that maximizes the transition temperature.   

We now evaluate the magnetic susceptibility of the FM and SGL phases in the weak-coupling limit by utilizing the
formalism developed by Fishman and Jarrell \cite{fis:03}.  Treating the electronic susceptibility as a 
$2\nc $-dimensional matrix, the local-moment susceptibility as a scalar, and the cross terms
as $2\nc $-dimensional vectors in Matsubara space,  
$\uchi (\vq =0,i\om )$ can be written as a $(2\nc  +1) \times (2\nc +1)$ supermatrix.  The total susceptibility
$\chi (\vq=0,i\om )$ is obtained by taking $\nc \rightarrow \infty $ and summing $\uchi (\vq =0, i\om )$ over all 
matrix elements for a fixed external frequency $\om =2m\pi T$.   As discussed in Ref.\cite{fis:03}, $\uchi (\vq =0, i\om )$ 
satisfies the Bethe-Salpeter equation 
\begin{equation}
\label{BS}
\uchi (\vq =0 ,i\om ) =\uchi^{(0)}(\vq =0 ,i\om )+\uchi^{(0)}(\vq =0 ,i\om ) \ugamma (i\om ) \uchi (\vq =0 ,i\om ),
\end{equation}
where $\ugamma (i\om )$ is the vertex function and $\uchi^{(0)}(\vq =0 ,i\om )$ is the
bare susceptibility.  Within DMFT, momentum conservation at the internal vertices of
irreducible graphs is disregarded so that internal Green's functions are replaced by their
local values.  Consequently, $\ugamma (i\om )$ is independent of momenta and may be evaluated
from an identical Bethe-Salpeter equation where $\underline{\chi }(\vq =0 ,i\om )$ and
$\underline{\chi}^{(0)}(\vq =0 ,i\om )$ are replaced by local susceptibilities at a site $i$.
Because the total spin $\sum_i (\vS_i +\vs_i)$ is conserved, the $\vq =0$ susceptibility $\chi (\vq=0,i\om )$
is proportional to $\delta_{m,0}$.  So we shall henceforth take $\om =0$ to evaluate the elastic
susceptibility $\chi (\vq=0)$. 

The magnetic field is taken to lie along the magnetization direction in the FM phase
but is averaged over all orientations in the SGL phase, which has no net magnetization.
For example, the bare local-moment susceptibility on site $i$ is given by $\beta S^2 \Bigl\{ 
\langle m_{i\alpha }^2\rangle -\langle m_{i\alpha }\rangle^2\Bigr\}
\Rightarrow \beta S^2 (1-q)/3$, where $q=M^2$ in the SGL phase and $q=3M^2-2(1-T/\TC) >0 $ ($T < \TC $)
or 0 ($T > \TC $) in the FM phase.  Following the derivation in Ref.\cite{fis:03}, we find that the total, elastic susceptibility
is given by
\begin{equation}
\label{sus}
\chi (\vq=0 ) = \frac{\Seff^2 }{3}\frac{1-q}{T-\TC (1-q)} + \frac{N_0(\mu )}{2},
\end{equation}
where $\Seff = S +\JH S N_0(\mu )$.  Since $N_0(\mu )\propto 1/W$, the electronic contribution $\JH S N_0(\mu )$
is much less than the local-moment contribution $S$ to $\Seff $ in the weak-coupling limit \cite{wc}.  
The electronic contribution enlarges or diminishes $\Seff $ depending on the sign
of $\JH $.  The final term in Eq.(\ref{sus}), $N_0(\mu )/2$, is just the electronic Pauli susceptibility.  
Since it does not depend on $\JH $ and does not diverge as $T\rightarrow \Tsg $, 
we shall neglect the Pauli susceptibility in the subsequent discussion.   For a FM, the
Curie-Wess susceptibility of Eq.(\ref{sus}) with $\Seff = S$ is precisely the same as the MF result for a
classical Heisenberg model \cite{yos:91}.   Bare in mind that the broad analogies between the (D)MF theories
of the DE and classical Heisenberg models only exist in the weak-coupling limit and disappear
once $\JH S$ becomes of order $W$.

Using the low-temperature behavior of $M(\tau )$, we find that $\qfm \rightarrow 1-\tau^2/3$ and 
$\chi (\vq=0)\rightarrow (\Seff^2/9)T/\TC^2 $ as $T\rightarrow 0$ in the FM phase. 
By contrast, $\qsg \rightarrow 1-2\tau /3$ and $\chi (\vq =0)\rightarrow (\Seff^2/3)/(3\Tsg /2-\TC )$ as 
$T\rightarrow 0$ in the SGL phase.  Because the SGL phase has no long-range order and the local moments 
have no preferred orientation for any $Q$ between 0 and 1, the zero-temperature susceptibility does not 
vanish as $p \rightarrow 0.26$ and $\Tsg \rightarrow \TC $.  The magnetic susceptibility in the SGL phase 
is plotted versus $\tau $ for several different concentrations in Fig.2.  As expected, the SGL susceptibility 
has a cusp at $\Tsg $, which develops into a divergence as $p\rightarrow 0.26$
and $Q\rightarrow 0$.    Notice that the normalized susceptibility $\Tsg \chi (\vq =0)/\Seff^2$
vanishes as $Q\rightarrow 1/2$ and $\Tsg \rightarrow 0$ in the vicinity of $p=0.5$.

\begin{figure}
\includegraphics *[scale=0.5]{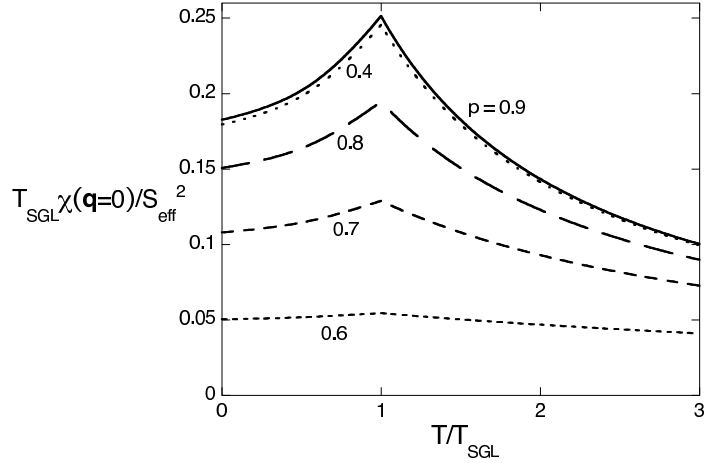}
\caption{
The elastic, $\vq =0 $ susceptibility versus $T/\Tsg $ for several electron concentrations $p$.
}
\end{figure}

Comparing Eq.(\ref{sus}) with the parameterization of Sherrington and Kirkpatrick (SK) \cite{sher:75}, 
we conclude that $q$ is the EA order parameter \cite{edw:75}.  We have plotted $\qsg $ and $\qfm $ versus $\tau $
in Fig.1.   Our result $\qsg = \langle \hm_i \rangle^2$ should be compared with the SK expression
$q=\langle \langle S_i\rangle^2\rangle_J$, where the inner expectation value is a thermal average for a given set of 
exchange couplings and the outer is an ensemble average over the distribution of exchange
couplings.  For the DE model, the EA order parameter is the same at every site even without an 
average over quenched disorder.  The DMFT result for the DE model differs from 
the SK result for the random $S=1/2$ Ising model in 
at least one important respect:   the SK prediction for $\sqrt{q}$ is {\it not} identical to the $S=1/2$ Brillouin function
whereas $\sqrt{\qsg }$ is equivalent to the $S=\infty $ Brillouin function.

To summarize, we have evaluated the total, magnetic susceptibility and EA order parameters
of the SGL phase in the weak-coupling or RKKY limit.  We find that the SGL susceptibility has a cusp at $\Tsg $
and reaches a nonzero constant as $T\rightarrow 0$, as expected for a phase with short-range
magnetic order.   The SGL phase is characterized by the short-range order of
$\langle \hm_i \rangle $ and by the nonzero value of $\langle \hm_i\rangle^2$
at every site.   These were the original SK criteria \cite{sher:75} for the existence of a SG.  
Unlike the SG ground state of the random Ising model, the SGL phase of the DE model 
can be studied analytically in the absence of quenched disorder.

It is a pleasure to acknowledge helpful conversations with Eugene Kogan and Thomas Maier. 
This research was sponsored by the U.S. Department of Energy under contract 
DE-AC05-00OR22725 with Oak Ridge National Laboratory, managed by UT-Battelle, LLC.


\begin{references}
\bibitem{ent}  The assumption of classical spins implies that for either the SGL or FM phases,
the specific heat goes to a constant and the entropy diverges as $T\rightarrow 0$, just as in the 
classical Heisenberg model.
\bibitem{wc}  Strictly speaking, only the local-moment $S^2$ term should be retained in the numerator of the
Curie-Weiss susceptibility in the weak-coupling limit.  However, to each order in $S$, Eq.(\ref{sus}) is formally 
correct to leading order in the dimensionless, small parameter $\JH S/W$ so long as quantum fluctuations of
the local-moments are neglected.



\suppressfloats
\end{references}
\end{document}